\documentclass[journal,onecolumn,draftclsnofoot,12pt]{IEEEtran}

\usepackage{pdfsync}
\usepackage{wrapfig}
\usepackage{multirow}
\usepackage{amsfonts}
\usepackage{amssymb}
\usepackage{amsmath}
\usepackage{graphicx}
\usepackage{pdfpages}
\usepackage{cite}
\usepackage[T1]{fontenc}
\usepackage{balance}
\usepackage{algorithm}
\usepackage{array}
\usepackage{algpseudocode}
\usepackage{float}
\usepackage{amsmath,amsthm,amssymb}
\usepackage{amsmath,amsthm}
\usepackage{bbm}
\usepackage{upgreek}

\usepackage{booktabs}
\usepackage{threeparttable}
\usepackage{subfigure}

\newcommand{\bs}{\boldsymbol}

\newtheorem{prop}{Proposition}

\DeclareMathOperator*{\argmin}{arg\,min}

\theoremstyle{lemma}

\begin{document}
\title{Asynchronous Massive Access in Multi-cell Wireless Networks Using Reed-Muller Codes}

\author{Pei~Yang,
        Dongning~Guo,
        and Hongwen Yang%
\thanks{Pei~Yang is with School of Wireless Communication Center, Beijing University of Posts and Telecommunications, Beijing 100876, China and also with the Department of Electrical and Computer Engineering, Northwestern University, Evanston, IL 60208 USA (e-mail: yp@bupt.edu.cn).}%
\thanks{Dongning~Guo is with the Department of Electrical and Computer Engineering, Northwestern University, Evanston, IL 60208 USA (e-mail: dguo@northwestern.edu).}
\thanks{Hongwen~Yang is with School of Wireless Communication Center, Beijing University of Posts and Telecommunications, Beijing 100876, China (e-mail: yanghong@bupt.edu.cn).}%
}


\maketitle

\begin{abstract}
\boldmath
Providing connectivity to a massive number of devices is a key challenge in 5G wireless systems. In particular, it is crucial to develop efficient methods for active device identification and message decoding in a multi-cell network with fading, path loss, and delay uncertainties.  This paper presents such a scheme using second-order Reed-Muller (RM) sequences and orthogonal frequency-division multiplexing (OFDM). For given positive integer $m$, a codebook is generated with up to $2^{m(m+3)/2}$ codewords of length $2^m$, where each codeword is a unique RM sequence determined by a matrix-vector pair with binary entries. This allows every device to send $m(m + 3)/2$ bits of information where an arbitrary number of these bits can be used to represent the identity of a node, and the remaining bits represent a message. There can be up to $2^{m(m+3)/2}$ different identities. Using an iterative algorithm, an access point can estimate the matrix-vector pairs of each nearby device, as long as not too many devices transmit in the same frame. It is shown that both the computational complexity and the error performance of the proposed algorithm exceed another state-of-the-art algorithm. The device identification and message decoding scheme developed in this work can serve as the basis for grant-free massive access for billions of devices with hundreds of simultaneously active devices in each cell.

\end{abstract}



\section{Introduction}
One of the promises of 5G wireless communication systems is to support large scale machine- type communication (MTC), that is, to provide connectivity to a massive number of devices as in the Internet of Things \cite{Durisi,Tullberg,Larsson}. In massive MTC scenarios, the connection density will be up to $10^6$ devices per square kilometer \cite{IMT}. A key characteristic of MTC is that the user traffic is typically sporadic so that in any given time interval, only a small fraction of devices are active. Also, short packets are the most common form of traffic generated by sensors and devices in MTC. This requires a fundamentally different design than that for supporting sustained high-rate mobile broadband communication.

A variety of effective multiple access techniques have been adopted in the previous and current cellular networks, such as frequency division multiple access (FDMA), time division multiple access (TDMA), code division multiple access (CDMA), and orthogonal frequency division multiple access (OFDMA) \cite{McCann}.
For these systems, resource blocks are orthogonally divided in time, frequency, or code domains.
This makes signal detection at the access point (AP) fairly simple because the interference between adjacent blocks is minimized.
However, due to the limitation of the number of orthogonal resource blocks, it can only support a limited number of devices.
To provide connectivity to a massive number of devices, an information-theoretic paradigm called \emph{many-user access} has been studied in \cite{Chen,Chen2}.
It was shown to be asymptotically optimal for active users to simultaneously transmit their identification signatures followed by their message bearing codewords.
Further, various massive access schemes have been proposed; see examples \cite{Yu,Yu2,Yu3,Hanzo,Zhang,Luo,Andreev,Polyanskiy,Poor,Ravi,Yener,Letaief,Sohrabi,Senel,Schober,Ahn,Lau,Jia,Howard,Robert,ThompsonCalderbank,ZhangLi,Applebaum,AndreevKowshik,Amalladinne,LiXiang,ChenGuo} and references therein.

Indeed, active device identification and channel estimation are initial steps to enable message decoding in MTC.
Due to the sporadic traffic in MTC, these problems are usually cast as neighbor discover or compressed sensing problems \cite{WZhu,Zhang,Luo,Amalladinne,ThompsonCalderbank,Yu,Yu2,Yu3,Sohrabi,Senel,Schober,Schepker,Gil,Giannakis,Jeong,Mir,Shim}.
When the channel coefficients are known at the AP, several compressed sensing schemes were proposed for active device identification \cite{Jeong,Mir,Shim}.
Further, approximate message passing (AMP)-based algorithms were applied for joint active device identification and channel estimation \cite{Sohrabi,Yu,Yu2,Yu3,Senel,Schober}.
In addition, greedy compressed sensing algorithm was designed for sparse signal recovery based on orthogonal matching pursuit \cite{Schepker,Gil}.


Another approach to the active device identification is slotted ALOHA.
Recently, an enhanced random access scheme called coded slotted ALOHA was proposed in \cite{Paolini,Casini}, where each message is repeatedly sent over multiple slots, and information is passed between slots to recover messages lost due to collision.
These works assume synchronized transmission and perfect interference cancelation.
The asynchronous model has been studied in \cite{Gaudenzi,Sandgren}.
It is pointed in \cite{Ordentlich} that slotted ALOHA only supports the detection of a single device within each slot.
Instead, the authors in \cite{Ordentlich} proposed the $T$-fold ALOHA scheme such that the decoder can simultaneously decode up to $T$ messages in the same slot.
By combining with serial interference cancellation, the performance of the $T$-fold ALOHA scheme is further improved in \cite{Vem}.  And \cite{Kowshik,AndreevKowshik} proposed the $T$-fold ALOHA based random access scheme for handing Rayleigh fading channels and asynchronous transmission.

To identify the active device from an enormous number of potential users in the system, each device must be assigned a unique sequence.
For a given positive integer $m$, a Reed-Muller (RM) code book is generated with up to $2^{m(m+3)/2}$ codewords of length $2^m$.
The code book size is so large that every user is assigned a different signature in any practical system.
Given this, RM sequence-based massive access schemes were proposed in \cite{Luo,Zhang,Robert,Howard,ZhangLi,Hanzo,YangGuo}.
In \cite{Howard}, a chirp detection algorithm for deterministic compressed sensing based on RM codes and associated functions was proposed.
The authors in \cite{Robert} have further enhanced the chirp detection algorithm with the slotting and patching framework.
However, the algorithm in \cite{Robert} works only for the additive white Gaussian noise (AWGN) channel (the channel estimation problem is thus not considered therein).
For fading channels, an iterative RM device identification and channel estimation algorithm is adopted in \cite{Hanzo} based on the derived nested structured of RM codes.
In our previous work \cite{YangGuo}, we extend the real codebook used in \cite{Hanzo} to the full codebook, which encodes more bits with little performance loss.
In addition, when the number of active devices is large, the performance of the algorithm in \cite{Hanzo} degrades dramatically.
In contrast, by adopting slotting and message passing, the algorithm in \cite{YangGuo} performs gracefully as the number of active devices increases.
It can be proved that the worst-case complexities of RM codes-based detection algorithms are sub-linear in the number of codewords, which makes it an attractive algorithm for message decoding in MTC.
The above RM detection algorithms are based on the assumption that the signals are synchronized.
However, due to propagation delays in the practical environment, asynchrony cannot be ignored.

In this paper, we investigate the joint device identification/decoding and channel estimation in an asynchronous setting. By removing the unrealistic assumption of fully synchronous transmissions, the proposed scheme is an important step towards a practical design.
In addition, 5G cellular systems are expected to deploy a large number of antennas to take advantage of massive multiple-input multiple-output (MIMO) technologies.
Massive MIMO has been extensively studied to enable massive connectivity \cite{Yu,Yu2,Yu3,Sohrabi,Senel,Schober,HHan,Carvalho,HRao}.
It can take advantage of the increased spatial degrees of freedom to support a large number of devices simultaneously.
Since the user traffic is sporadic in MTC, \cite{Yu,Yu2,Yu3,Sohrabi,Senel,Schober} proposes to formulate the device identification problem based on compressed sensing and thereby can be solved by the computationally efficient AMP algorithm.
And a new pilot random access protocol called strongest-user collision resolution is proposed in \cite{HHan,Carvalho} to solve the intra-cell pilot collision in crowded massive MIMO systems.
We note that algorithms using random codes and/or AMP type of decoding do not scale to millions of potential devices.
Moreover, the preceding algorithms are based on the assumption that the signals are synchronized.
{As far as we know, there has been no research publications on asynchronous massive access with a large number of users and antennas.}  To fill this gap, we investigate asynchronous massive access in a multi-cell wireless network using Reed-Muller codes with many receive antennas which has the potential to support billions of potential devices in the network.


The main contributions are summarized as follows:
\begin{itemize}
\item Compared with the algorithms in \cite{Robert,Howard,ZhangLi,Hanzo,YangGuo} where transmitted signals are synchronized, we extend the algorithms to an asynchronous case where the arbitrary delays of each device are estimated based on the derived relationship between an RM sequence and its subsequences.
\item To enhance the performance, we extend the RM detection algorithms to the case where the AP is deployed with a large number of antennas.
\item We further describe an enhanced RM coding scheme with slotting and bit partition. The corresponding detection algorithm is referred to as Algorithm 1. We show that the computational complexity and performance of Algorithm 1 are notably improved, which makes it one important step closer to a practical algorithm.
\item While many papers in the literature study massive access, this work is one of the few that can truly accommodate billions of devices and more.
\end{itemize}

The remainder of this paper is organized as follows.
The system model is presented in Section \ref{Secsystemmodel}.
Section \ref{RMrelationship} outlines the relationship between the RM sequence and its subsequences, which is the basis of the RM asynchronous decoding algorithm.
Section \ref{useridentification} displays the enhanced RM decoding algorithm utilizing slotting, message passing, and bit partition.
Further, the computation complexity analysis is given in Section \ref{CCA}.
Section \ref{Numericalresults} presents the numerical results, while Section \ref{conclusion} concludes the paper.

Throughout the paper, boldface uppercase letters stand for matrices while boldface lowercase letters represent column vectors.
The superscripts $(\cdot)^{\rm T}$, $(\cdot)^{*}$, and $(\cdot)^\dag$ denote the transpose, complex conjugate, and conjugate transpose operator, respectively.
The complex number field is denoted by $\mathbb C$.
$\|\bs x\|_p$ denotes the $p$-norm of a vector $\bs x$, ${\|\bs X\|}_F$ denotes the Frobenius norm of a matrix $\bs X$, and $|A|$ denotes the cardinality of set $A$.
$\bs I_n$ denotes an $n\times n$ identity matrix.
$\lceil x\rceil$ represents the rounding function, which returns the smallest
integer greater than $x$.
$\odot$ means element-wise multiplication and ${\rm Arg}(\cdot)\in[-\pi, \pi)$ gives the phase angle of a complex number.


\section{System Model}\label{Secsystemmodel}

\subsection{Transmission Scheme}
Let $\Phi$ denote a large but finite set of devices on the plane with area $S$, where each device is equipped with one antenna.
Further we denote ${\cal K}\subseteq \Phi$ as the active device set on the plane, each of which has $B$ bits to be sent.
Due to the sporadic traffic in MTC, the number of active devices in a given time interval is far less than the total number of devices, i.e., $|{\cal K}|\ll |\Phi|$.

Prior to transmission, a message of $B$ bits is partitioned into $2^d$ sub-blocks, where the $j$-th sub-block consists $B_j$ information bits such that $\sum_{j=1}^{2^d}B_j=B$.
To patch the information bits in different sub-blocks together, we adopt the tree encoder proposed in \cite{Amalladinne2}.
Specifically, the tree encoder appends $l_j$ parity bits to sub-block $j$, where the appended check bits satisfy random parity constraints associated with the message bits contained in the previous sub-blocks ($l_1=0$ in all cases).
These parity check bits are needed to patch the information bits in different patches together, but they do not serve the purpose of transmitting the information.
All the sub-blocks have the same size, i.e., $B_j+l_j=J$.

Assume there are $2^p$ time slots. Each active device randomly selects 2 slots to send its $J$ bits.
We use $p$ bits to encode the location of the primary slot.
And we use an arbitrary subset of size $p$ to encode a \emph{translate}, which gives the secondary slot location when it is added to the primary slot location.
To distinguish the primary and secondary slots, we fix a single \emph{check bit} in the information bits to be 0 for the primary slot and 1 for the secondary slot. Thus deducing 1 bit from the total number of bits transmitted.
Besides, in this paper, we deal with asynchronous transmission.
To estimate the device delay, we assume two information bits to be zeros (To be specified in Section \ref{findpb}).

\begin{figure}
  \begin{center}
  \includegraphics[width=6in]{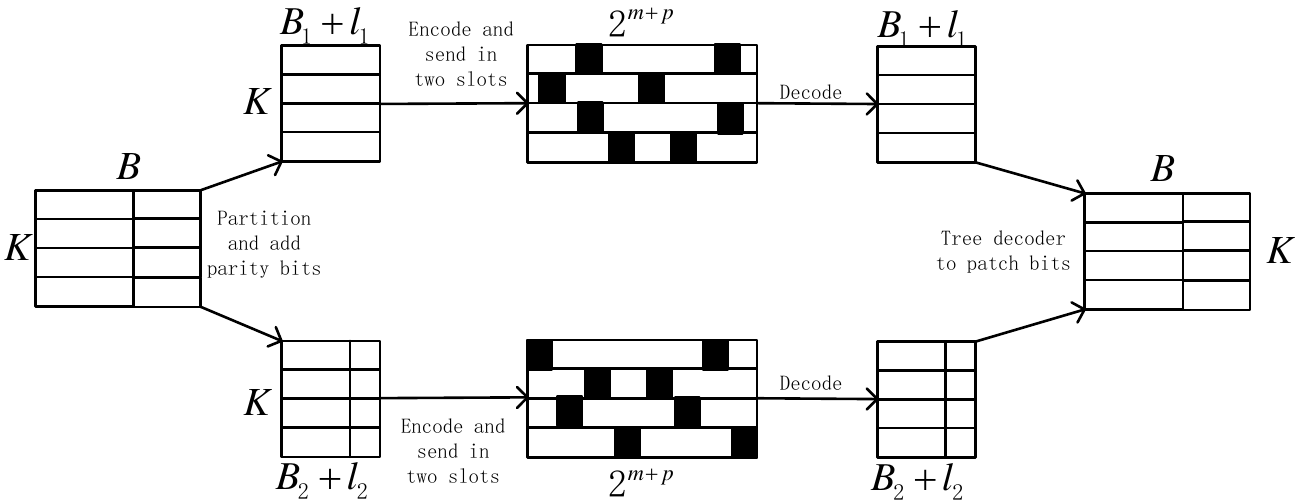}
  \end{center}
  \caption{Illustration of the transmission scheme. We set $d=1$ and $K=4$ for simplicity.}\label{sottingandpatching}
\end{figure}

Fig. \ref{sottingandpatching} depicts the transmission scheme where we set $d=1$ and $K=4$ for simplicity. In Fig. \ref{sottingandpatching}, each device has $B$ information bits to be sent. The information bits is first divided into $2^d$ sub-blocks. Each sub-block contains $B_j$ information bits, along with $l_j$ parity bits such that $J=B_j+l_j$. Then each device sent 2 copies of the $J$ bits in 2 randomly selected slots within the $2^p$ slots.
Finally we perform the proposed decoding scheme and use tree decoder to patch together the information in different sub-blocks.

\subsection{Encoding}

Our approach is to encode these $J$ bits in each slot to a length $N=2^m$ second-order RM codes.
A length $2^m$ second-order RM sequence is determined by a symmetric binary matrix $\bs P^m\in{\mathbb Z}_2^{m\times m}$ and a binary vector $\bs b^m\in{\mathbb Z}_2^{m}$.
Since $\bs P^m$ is determined by $\frac12m(m+1)$ bits and $\bs b^m$ is determined by $m$ bits, each sequence encodes $\frac12m(m+3)$ bits.
Given the matrix-vector pair $(\bs P^m,\bs b^m)$, the $n$-th entry of the RM sequence $\bs X^m$ can be written as \cite{Robert}
\begin{equation}
\begin{split}\label{cm2}
    X^m_n={\iota}^{ 2(\bs b^m)^{\rm T}\bs a_{n-1}^m+(\bs a_{n-1}^m)^{\rm T}\bs P^m\bs a_{n-1}^m },\quad n=1,\cdots,N,
\end{split}
\end{equation}
where $\iota^2=-1$, $\bs a_{n-1}^m$ is the $m$-bit binary expression of $(n-1)$. Eq. \eqref{cm2} indicates that $X^m_n\in \{1,-1,\iota,-\iota\}$.

In this case, we have
\begin{align}
    J=\frac12m(m+1)+p-3.
\end{align}
Further, the number of information bits is written as
\begin{align}\label{B}
    B=2^d\left(\frac12m(m+3)+p-3\right)-\sum\limits_{j=1}^{2^d}l_j
\end{align}


\subsection{Channel Model}

In this paper, we consider OFDM modulation where each symbol consists of $N$ subcarriers.
Denote the frequency samples of device $k$ as $X_{k,n}^m$ where $n=1,2,\cdots,N$ is the subcarrier index.
As explained before, $X_{k,n}^m$ is a RM sequence generated by \eqref{cm2}.
The time-domain OFDM symbol of device $k$ can be written as
\begin{align}\label{OFDMmodel}
    x_k(t)=\sqrt{\gamma}\sum_{n=1}^NX_{k,n}^m{\rm e}^{2\pi \iota\Delta fnt},\quad t\in\left[0,\frac1{\Delta f}+\tau_{\rm max}\right],
\end{align}
where $\Delta f$ is the carrier spacing and the symbol duration is $1/\Delta f$ and $\tau_{\rm max}$ is the maximum device delay. $X_{k,n}^m$ is device $k$'s sample to be transmitted in subcarrier $n$. $\gamma$ denotes the transmit power.

We denote ${\cal K}_i$ as the active device set that transmitted in time slot $i$.
Let $K=|{\cal K}|,K_i=|{\cal K}_i|$, we have $K_i\approx 2K/2^p$ since each device randomly choose 2 slots in the $2^p$ time slots.
Without loss of generality, we assume the index of the active devices in slot $i$ is ${\cal K}_i=\{1, 2, \cdots, K_i\}$.
We focus on one AP equipped with $r$ antennas, and assume that the AP is located at the origin of the plane.
The receive signal of the $l$-th antenna of the AP at time slot $i$ is written as
\begin{align}\label{OFDMAPsignal}
    y_{l,i}(t)&=\sum_{k=1}^{K_i}h_{k,l}x_k(t-\tau_k)+z_{l,i}(t),
\end{align}
where $\bs h_k=[h_{k,1},\cdots,h_{k,r}]^{\rm T}$ is the channel vector between device $k$ and the AP and $z_l(t)$ is additive white Gaussian noise; $\tau_k$ is the transmission delay and $x_k(t)$ is the transmit signal of device $k$.

Then the AP samples at time $\frac1{\Delta f}\frac{u}{N},u=1,\cdots,N+\left\lceil\tau_{\rm max}N\Delta f\right\rceil$. The total number of samples in each slot is thus $N+M$ where $M=\left\lceil\tau_{\rm max}N\Delta f\right\rceil$ is the length of cyclic prefix.
Furthermore, the total codelength can be written as
\begin{align}\label{Codelength}
    C=2^{d+p}(N+M).
\end{align}
Then the AP discard the first $M$ cyclic prefix in the OFDM symbol to form the discrete-time receive signal
\begin{align}\label{OFDMAPdiscrete}
    y_{l,i}(u)=\sqrt{\gamma}\sum_{k=1}^{K_i}h_{k,l}\sum_{n=1}^NX_{k,n}^m{\rm e}^{- \iota\Delta_k n}{\rm e}^{\iota\frac{2\pi }{N}nu}+z_{l,i}(u),\quad u=1,\cdots,N,
\end{align}
where $\Delta_k=2\pi\Delta f\tau_k$ is the normalized delay; $z_{l,i}(u)\sim{\cal CN}(0,1)$. We assume the normalized delay is uniformly distributed in $\Delta_k\in[-\pi,\pi]$.

Performing $N$ point DFT on $[y_{l,i}(1),\cdots,y_{l,i}(N)]^{\rm T}$ yields
\begin{align}
    Y_{l,i}^m(n)&=\sqrt{\gamma}\sum_{k=1}^{K_i}h_{k,l}\frac1{N}\sum_{u=1}^N{\rm e}^{-\iota\frac{2\pi }{N}nu}\sum_{v=1}^NX_{k,v}{\rm e}^{- \iota\Delta_k v}{\rm e}^{\iota\frac{2\pi }{N}uv}+Z_{l,n}^m(u)\\
    &=\sqrt{\gamma}\sum_{k=1}^{K_i}h_{k,l}X_{k,n}{\rm e}^{-\iota\Delta_kn}+Z_{l,i}^m(n)\label{OFDMAPdiscreteDFT},
\end{align}
where
\begin{align}\label{OFDMAPdiscreteDFTnoise}
    Z_{l,i}^m(n)&=\frac1{N}\sum_{u=1}^N{\rm e}^{-\iota\frac{2\pi }{N}nu} z_{l,i}(u)\sim{\cal CN}(0,1).
\end{align}
and $n=1,\cdots,N$ and $l=1,\cdots, r$.

Let $\bs Y_{i}^m(n)=[Y_{1,i}^m(n),\cdots,Y_{r,i}^m(n)]^{\rm T}$. For simplicity, denote $\bs Y_i^m=\left[\bs Y_{i}^m(1),\cdots,\bs Y_{i}^m(N)\right]$ as the DFT results in slot $i$.

\subsection{Propagation Model and Cell Coverage}\label{PM}
Consider a multiaccess channel with active devices distributed across the plane according to a homogeneous Poisson point process with
intensity $\lambda$. The number of active devices on the plane with its area equal to $S$ is a Poisson random variable with mean $\lambda S\approx |\cal K|$.

We further divide the active device set ${\cal K}$ into in-cell device set (neighbor) and out-of-the-cell device set (non-neighbor) of the AP according to the nominal SNR between the AP and the devices. If the nominal SNR between a device and the AP is larger than a threshold, then this device is considered an in-cell device of the AP.
The purpose of the AP is to identify all in-cell devices and/or decode their messages, where transmissions from out-of-the-cell devices are regarded as interference.

The small-scale fading between the device and the AP is modeled by an independent Rayleigh random variable with unit mean.
The large-scale fading is modeled by the free-space path loss which attenuates over distance with some path loss exponent $\alpha>2$.

Let $D_{k}$ and $\bs G_k=[G_{k,1},\cdots,G_{k,r}]^{\rm T}$ denotes the distance and the small scale Rayleigh fading gain between device $k, k=1,2,\cdots,K$, and the AP, respectively.
Then the channel gain between device $k$ and the $l$-th antenna of the AP is expressed as
\begin{equation}
\begin{split}
    |h_{k,l}|^2=D_k^{-\alpha}G_{k,l},
\end{split}
\end{equation}
where the phase of $h_{k,l}$ is uniformly distributed on $[0,2\pi)$.

The coverage of the AP can be defined in many different ways. According to \cite{ZhangGuo}, device $k$ and the AP are neighbors of each other if the channel gain exceeds a certain threshold $\theta$.
Assume device $k$ and the AP are neighbors, i.e., $D_k^{-\alpha}\|\bs G_k\|_1>r\theta$, we have $D_k<\left({\frac{\|\bs G_k\|_1}{r\theta}}\right)^{1/{\alpha}}$.
Under the assumption that all devices form a p.p.p., for given $\bs G_k$, device $k$ is uniformly distributed in a disk centered at the AP with radius $\left({\frac{\|\bs G_k\|_1}{r\theta}}\right)^{1/{\alpha}}$.
The average number of neighbors of the AP is calculated as
\begin{align}\label{averageneighbor}
    K^{*}&=\mathbb E_{\bs \Phi}\left\{\sum\limits_{k\in\bs\Phi}1(D_k^{-\alpha}\|\bs G_k\|_1\ge r\theta)\right\}\\
         &=2\pi\lambda\int_0^{\infty}\int_0^{\infty}1(gs^{-\alpha}\ge r\theta)s\frac1{\Gamma(r)}g^{r-1}{\rm e}^{-g}{\rm d}s{\rm d}g\\
         &=2\pi\lambda\int_0^{\infty}\int_0^{\left(\frac{g}{r\theta}\right)^\frac1{\alpha}}s\frac1{\Gamma(r)}g^{r-1}{\rm e}^{-g}{\rm d}s{\rm d}g\\
         &=\pi\lambda\int_0^{\infty}\left(\frac{g}{r\theta}\right)^\frac2{\alpha}\frac1{\Gamma(r)}g^{r-1}{\rm e}^{-g}{\rm d}g\\
         &=\pi\lambda(r\theta)^{-\frac2{\alpha}}\frac{\Gamma\left(\frac2{\alpha}+r\right)}{\Gamma(r)}\label{Kstar}
\end{align}
where $\Gamma(\cdot)$ is the Gamma function and $1(\cdot)$ is the indicator function.
Eq. \eqref{Kstar} indicates that $K^{*}$ is an increasing function of $r$.

In addition, the sum power of all out-of-the-cell devices can be derived as
\begin{align}\label{nonneighbor}
    \sigma^2&=\mathbb E_{\bs \Phi}\left\{\sum\limits_{k\in\bs\Phi}1(D_k^{-\alpha}\|\bs G_k\|_1< r\theta)\gamma D_k^{-\alpha}\|\bs G_k\|_1\right\}\\
    &=2\pi\lambda\gamma\int_0^{\infty}\int_0^{\infty}gs^{-\alpha}1(gr^{-\alpha}< r\theta)s\frac1{\Gamma(r)}g^{r-1}{\rm e}^{-g}{\rm d}s{\rm d}g\\
    &=2\pi\lambda\gamma\int_0^{\infty}\int_{\left(\frac{g}{r
    \theta}\right)^\frac1{\alpha}}^{\infty}s^{1-\alpha}\frac1{\Gamma(r)}g^{r}{\rm e}^{-g}{\rm d}s{\rm d}g\\
    &=\frac{2\pi\lambda\gamma}{\alpha-2}\int_0^{\infty}\left(\frac{g}{r\theta}\right)^\frac{2-\alpha}{\alpha}\frac1{\Gamma(r)}g^{r}{\rm e}^{-g}{\rm d}g\\
    &=(r\theta)^{1-2/\alpha}\frac{2\pi\lambda\gamma}{(\alpha-2)}\frac{\Gamma\left(\frac2{\alpha}+r\right)}{\Gamma(r)}
\end{align}

\section{A Property of RM Sequences}\label{RMrelationship}
Before given the decoding algorithm, we first derive a property of RM sequence, which is the basis of our decoding algorithm.

Let $m$ be a given positive number. Let $\bs b^s=[b^m_1,b_2^m,\cdots,b_s^m]^{\rm T}$ be a binary $s$-tuple. For $s=2,\cdots,m$, we have
\begin{equation}
\begin{split}\label{b}
    \bs b^s=\left[\begin{array}{c}
               \bs b^{s-1} \\
               b_s^m
             \end{array}\right].
\end{split}
\end{equation}
Furthermore, let $P^1=[\beta^m_1]$. For $s=2,\cdots,m$, let the $s\times s$ binary matrix $\bs P^s$ be defined recursively as
\begin{equation}
\begin{split}\label{P}
\bs P^s=\left[\begin{array}{cccc}
  \bs P^{s-1} & \bs \eta^s \\
  (\bs \eta^s)^{\rm T} & \beta^m_s
\end{array}\right]
\end{split},
\end{equation}
where $[\beta^m_1,\beta_2^m,\cdots,\beta_s^m]^{\rm T}$ is the main diagonal elements of $\bs P^s$, and $\bs\eta^s$ is a length $s-1$ column vector.

We have the following result.
\begin{prop}\label{Proposition1}
Given a length-$2^m$ RM sequence, its order $s$ and $s-1$ sub-sequences satisfy
\begin{equation}
\begin{split}\label{structure}
    \left\{\begin{array}{l}
      X^s_{2n}=V_n^{s-1}X_n^{s-1} \\
      X^s_{2n-1}=X_n^{s-1}
    \end{array}\right.,\quad n=1,\cdots,2^{s-1}, s=2,\cdots,m
\end{split}
\end{equation}
where
\begin{align}\label{v}
V_n^{s-1}&=(-1)^{b_s^m+\frac12\beta_s^m+(\bs \eta^s)^{\rm T}\bs a_{n-1}^{s-1}}.
\end{align}
The vector $[V_1^{s-1},\cdots,V_{2^{s-1}}^{s-1}]^{\rm T}$ is a length-$2^{s-1}$ Walsh sequence with frequency $\bs\eta^s$.

\end{prop}

\begin{IEEEproof}
Recall $a_n^s$ is the $s$-bit expression of $n$. For $n=1,\cdots,2^{s-1}$, the vector $\bs a_{2n-1}^s$ can be decomposed as
\begin{equation}
\begin{split}
    \bs a_{2n-1}^s=\left[\begin{array}{c}
               \bs a_{n-1}^{s-1} \\
               1
             \end{array}\right].
\end{split}
\end{equation}
Consequently
\begin{align}
    &2(\bs b^s)^{\rm T}\bs a_{2n-1}^s+(\bs a_{2n-1}^s)^{\rm T}\bs P^s\bs a_{2n-1}^s\notag\\
&= \left[\begin{array}{cc}
                                         \left(\bs a_{n-1}^{s-1}\right)^{\rm T} & 1
                                       \end{array}\right]
           \left[\begin{array}{cccc}
  \bs P^{s-1} & \bs \eta^s \\
  (\bs \eta^s)^{\rm T} & \beta^m_s
\end{array}\right]\left[\begin{array}{c}
               \bs a_{n-1}^{s-1} \\   1
             \end{array}\right]+2(\bs b^s)^{\rm T}\left[\begin{array}{c}
               \bs a_{n-1}^{s-1} \\   1
             \end{array}\right]\\\label{2j1}
&=2(\bs b^{s-1})^{\rm T}\bs a_{n-1}^{s-1}\!+2b_s^m\!+\beta_s^m+\!2(\bs \eta^s)^{\rm T}\bs a_{n-1}^{s-1}\!+\left(\bs a_{n-1}^{s-1}\right)^{\rm T}\!\bs P^{s-1}\bs a_{n-1}^{s-1}.
\end{align}
Substituting \eqref{2j1} into \eqref{cm2} yields
\begin{align}
X^s_{2n}&=X_n^{s-1}\iota^{2b_s^m+\beta_s^m+2(\bs \eta^s)^{\rm T}\bs a_{n-1}^{s-1}}\\
&=V_n^{s-1}X_n^{s-1}.
\end{align}

Likewise, the binary vector $\bs a_{2j-2}^s$ can be decomposed as
\begin{equation}
\begin{split}
    \bs a_{2n-2}^s=\left[\begin{array}{c}
               \bs a_{n-1}^{s-1} \\
               0
             \end{array}\right]
\end{split}.
\end{equation}
Then the exponent of $X^s_{2n-1}$ is expressed as
\begin{align}
    &2(\bs b^s)^{\rm T}\bs a_{2n-2}^s+(\bs a_{2n-2}^s)^{\rm T}\bs P^s\bs a_{2n-2}^s\notag\\
&=\left[\begin{array}{cc}
                                         \left(\bs a_{n-1}^{s-1}\right)^{\rm T} & 0
                                       \end{array}\right]
           \left[\begin{array}{cccc}
  \bs P^{s-1} & \bs \eta^s \\
  (\bs \eta^s)^{\rm T} & \beta_s^m
\end{array}\right]\left[\begin{array}{c}
               \bs a_{n-1}^{s-1} \\   0
             \end{array}\right]+ 2(\bs b^s)^{\rm T}\left[\begin{array}{c}
               \bs a_{n-1}^{s-1} \\  0
             \end{array}\right]\\\label{2j}
&=2(\bs b^{s-1})^{\rm T}\bs a_{n-1}^{s-1}+\left(\bs a_{n-1}^{s-1}\right)^{\rm T}\bs P^{s-1}\bs a_{n-1}^{s-1}.
\end{align}
Substituting \eqref{2j} into \eqref{cm2} yields
\begin{equation}
\begin{split}
    X^s_{2n-1}=X_n^{s-1}.
\end{split}
\end{equation}
\end{IEEEproof}
{\emph {Remark 1}:} The structure of the derived RM sequence is similar to that of given in \cite{Hanzo}. The differences are two folds: 1) given the code length $2^m$, compared with the structure given in \cite{Hanzo}, the structure given in this paper allows us to send $2m$ more bits of information; 2) the way we splitting the sequences is different.

\section{Device Identification/Decoding and Channel Estimation}\label{useridentification}
In this section, we propose a novel RM asynchronous detection algorithm for active device detection and channel estimation that leverages Proposition \ref{Proposition1}.

\subsection{RM Asynchronous Detection Algorithm}
According to Fig. \ref{sottingandpatching}, the AP decodes the messages in different sub-blocks in a sequential manner.
In each sub-block, the AP decodes the messages slot-by-slot. Since each device transmits in 2 time slots, the message decoded in the previous time slot will be propagated to another time slot to eliminate its interference.

The detailed algorithm is summarized as in Algorithm 1.

\begin{center}
\begin{tabular}{l l}
\toprule
\multicolumn{1}{l}{{\bf Algorithm 1}: RM asynchronous detection algorithm.}\\
\midrule
{\bf Input}: the received signal $[\bs Y^q_1,\cdots,\bs Y^q_{2^p}]$, the average number of devices $K_{\rm max}$ in each\\
slot.\\
{\bf for} $patch=1:2^d$ {\bf do}\\
\quad Set $\bs P=[\ ]$, $\bs b=[\ ]$, $slot=[\ ], \bs h=[\ ]$, $\bs \Delta=[\ ]$, $t=0$.\\
\quad{\bf for} $i=1:2^p$ {\bf do}\\
\quad\quad $k\leftarrow0$.\\
\quad\quad {\bf for }$j=1:s$ {\bf do}\\
\quad\quad\quad {\bf if} $slot[j]=i$ {\bf do}\\
\quad\quad\quad\quad Remove the interference of device $j$ in slot $i$ and update $\bs Y^q_i$ according to \eqref{OFDMAPdiscreteDFTSIC}.\\
\quad\quad\quad\quad $t\leftarrow t+1$.\\
\quad\quad\quad {\bf end if}\\
\quad\quad {\bf end for}\\
\quad\quad $(\hat{\bs P}^m,\hat{\bs b}^m,\hat{\bs h}, \hat{\Delta})\leftarrow {\bf findPb}\ (\bs Y_i^q)$.\\
\quad\quad Denote $k_1$ as the number of detected messages in slot $i$.\\
\quad\quad {\bf for} $j=1:k_1$\\
\quad\quad\quad {\bf if} $(\hat{\bs P}_j^m, \hat{\bs b}_j^q)$ are not recorded in $(\bs P, \bs b)$ {\bf do}\\
\quad\quad\quad\quad $t\leftarrow t+1$.\\
\quad\quad\quad\quad $\bs P[:,:,t]\leftarrow \hat{\bs P}_{j}^m$.\\
\quad\quad\quad\quad $\bs b[:,t]\leftarrow \hat{\bs b}_{j}^m$.\\
\quad\quad\quad\quad $\bs h[:,t]\leftarrow \hat{\bs h}_{j}$.\\
\quad\quad\quad\quad $\bs\Delta[t]\leftarrow \hat{\Delta}_{j}$.\\
\quad\quad\quad\quad Calculate the translate according to $\left(\hat{\bs P}_j^q, \hat{\bs b}_j^q\right)$ and update $slot[s]$.\\
\quad\quad\quad {\bf end if }\\
\quad\quad {\bf end for}\\

\quad{\bf end for}\\
\quad Record $\bs P, \bs b, \bs h$, and $\bs\Delta$ in each sub-block.\\
{\bf end for}\\
{\bf Output}: Using tree decoder to patch the information bits together and output.\\
\bottomrule
\end{tabular}
\end{center}

The {\bf findPb} algorithm in Algorithm 1 returns all the messages transmitted in slot $i$, including the information bits $(\bs P, \bs b)$, the channel vector $\bs h$, and the device delay $\Delta$.
The {\bf findPb} algorithm decodes the messages transmitted in slot $i$ in a sequential manner.
Assume the channel gain of device $k\in\{1,\cdots,K_i\}$ is the biggest.
We will show that device $k$ can be first estimated from the received signal \eqref{OFDMAPdiscreteDFT}.
After device $k$ is detected, the AP performs successive interference cancellation (SIC) to remove the interference of device $k$ to detect the remaining devices.
This requires the AP to estimate not only the matrix-vector pair $(\bs P_k^m,\bs b_k^m)$, but also the device delay $\Delta_k$.
In this paper, to estimate the device delay, we let
\begin{align}
    b_{l,m}^m=\beta_{l,m}^m=0, \quad l=1,\cdots,K_i.
\end{align}
For simplicity, let $\bs \Delta_k^m=[\Delta_{k,1}^m, \cdots, \Delta_{k,m}^m]$, where
\begin{align}
    \Delta_{k,l}^m={\rm Arg}\left({\rm e}^{2^{l-1}j\Delta_k}\right),\quad l=1,\cdots, m.
\end{align}

In the next section, we show how to estimate the messages of device $k$.

\subsection{The {\bf findPb} Algorithm}\label{findpb}
According to \eqref{b} and \eqref{P}, the matrix-vector pair $(\bs P^m,\bs b^m)$ is determined by $(\bs\eta^{s},b_s^m,\beta_s^m), s=m,\cdots,2$ and $(b_{1}^m,\beta_{1}^m)$.
Specifically, the matrix-vector pair of the $k$th device $(\bs P_k^m,\bs b_k^m)$ will be estimated recursively.
We will show that the algorithm first estimates $\bs \eta_k^{m}$, then $(\bs \eta_k^{m-1},b_{k,m-1}^m,\beta_{k,m-1}^m)$, and finally the channel coefficient $\bs h_k$, $(b_{k,1}^m,\beta_{k,1}^m)$, and ${\Delta}_k$.

Then the receiver signal in slot $i$ is updated as
\begin{align}
    \bs Y_{i}^m(n)\leftarrow \bs Y_{i}^m(n)-\sqrt{\gamma}\bs h_{k}X_{k,n}^m{\rm e}^{-j\Delta_{k}n}\label{OFDMAPdiscreteDFTSIC},
\end{align}
for detecting the remaining devices.


\subsubsection{Estimation of $(\bs \eta_k^{m},\Delta_{k,1}^m)$}\label{sectionalpham}
From \eqref{OFDMAPdiscreteDFT} and \eqref{structure}, when $n=1,\cdots,2^{m-1}$, we have
\begin{align}
    \bs Y_{i}^m(2n)&=\sqrt{\gamma}\sum_{k=1}^{K_i} \bs h_k X_{k,2n}^m{\rm e}^{-2j\Delta_kn}+\bs Z_{i}^m(2n)\\
    &=\sqrt{\gamma}\sum_{k=1}^{K_i} \bs h_k V_{k,n}^{m-1}X_{k,n}^{m-1}{\rm e}^{-2j\Delta_kn}+\bs Z_{i}^m(2n)\label{y2j0},
\end{align}
and
\begin{align}
    \bs Y_{i}^m(2n-1)&=\sqrt{\gamma}\sum_{k=1}^{K_i}\bs  h_k X_{k,2n-1}^m{\rm e}^{-j\Delta_k(2n-1)}+\bs Z_{i}^m(2n-1)\\
    &=\sqrt{\gamma}\sum_{k=1}^{K_i} \bs h_k X_{k,n}^{m-1}{\rm e}^{-j\Delta_k(2n-1)}+\bs Z_{i}^m(2n-1)\label{y2j1}.
\end{align}

Define $\tilde Y_n^{m-1}=[\bs Y_{i}^m(2n)]^{\rm T}[\bs Y_{i}^m(2n-1)]^{*}$ for $n=1,\cdots,2^{m-1}$, \eqref{y2j0} and \eqref{y2j1} lead to
\begin{align}
    \tilde Y_n^{m-1}&=\gamma\sum_{k=1}^{K_i} \left\|\bs h_kX_{k,n}^{m-1}\right\|^2 V_{k,n}^{m-1}{\rm e}^{-j\Delta_k} +  \tilde {Z}_{n}^{m-1}\\
    &=\gamma\sum_{k=1}^{K_i} \|\bs h_k\|^2 V_{k,n}^{m-1}{\rm e}^{-j\Delta_k}+  \tilde Z_{n}^{m-1}\label{ym1},
\end{align}
where
\begin{equation}
\begin{split}
    \tilde {Z}_{n}^{m-1}&= \gamma\sum_{l=1}^{K_i}\sum_{k\ne l} \bs h_k^{\rm T}V_{k,n}^{m-1}X_{k,n}^{m-1}(\bs h_lX_{l,n}^{m-1})^{*}{\rm e}^{-2j\Delta_kn}{\rm e}^{j\Delta_l(2n-1)}\\
    &+\sqrt{\gamma}\sum_{k=1}^{K_i} \bs h_k^{\rm T} V_{k,n}^{m-1}X_{k,n}^{m-1}{\rm e}^{-2j\Delta_kn}(\bs Z_{i}^m(2n-1))^{*}\\
    &+\sqrt{\gamma}(\bs Z_{i}^m(2n))^{\rm T}\sum_{l=1}^{K_i} (\bs h_lX_{l,n}^{m-1}{\rm e}^{-j\Delta_k(2n-1)})^{*} +(\bs Z_{i}^m(2n))^{\rm T}(\bs Z_{i}^m(2n-1))^{*}.
\end{split}
\end{equation}
The first term in the right-hand side of \eqref{ym1} is a linear combination of Walsh functions $V_{k,n}^{m-1}, k=1,2,\cdots,K$, with frequency $\bs\eta^m_k$, which can be recovered by applying Walsh-Hadamard Transformation (WHT). The second term, $\tilde {Z}_{n}^{m-1}$, is a linear combination of chirps, which can be considered to be distributed across all Walsh functions to equal degree, and therefore these cross-terms appear as a uniform noise floor.

Let the Hadamard matrix be $\bs W^m=\left[\bs w_1^m,\bs w_2^m,\cdots, \bs w_{2^m}^m\right]^{\rm T}$ and its $(l,n)$-th elements are $\bs W_{l,n}^m=(-1)^{(\bs a_{l-1}^m)^{\rm T}{\bs a_{n-1}^m}}$. Denote the WHT transformation as $\bs t^{m-1}=\bs W^{m-1}\tilde {\bs Y}^{m-1}$, where $\tilde {\bs Y}^{m-1}=[\tilde{Y}_1^{m-1},\cdots,\tilde{Y}_{2^{m-1}}^{m-1}]$. The $l$-th entry of $\bs t^{m-1}$ can be written as
\begin{align}
    t_l^{m-1}&=(\bs w_l^{m-1})^{\rm T}\tilde {\bs Y}^{m-1}\\
    &=\sum_{n=1}^{2^{m-1}}(-1)^{(\bs a_{l-1}^{m-1})^{\rm T}{\bs a_{n-1}^{m-1}}}\left(\gamma\sum_{k=1}^{K_i} \|\bs h_k\|^2 V_{k,n}^{m-1}{\rm e}^{-j\Delta_k} \!+\!  \tilde {Z}_{n}^{m-1}\right)\\
    &=\gamma\sum_{n=1}^{2^{m-1}}(-1)^{\left(\bs a_{l-1}^{m-1}\right)^{\rm T}{\bs a_{j-1}^{m-1}}}\sum_{k=1}^{K_i}{\rm e}^{-j\Delta_k} \|\bs h_k\|^2 (-1)^{b_{k,m}^m+\frac12\beta_{k,m}^m+(\bs \eta_k^m)^{\rm T}\bs a_{n-1}^{m-1}}\notag\\
    &+\sum_{n=1}^{2^{m-1}}(-1)^{(\bs a_{l-1}^{m-1})^{\rm T}{\bs a_{n-1}^{m-1}}}\tilde {Z}_{n}^{m-1}\label{hadaym0}
\end{align}
Equation \eqref{hadaym0} can be further written as
\begin{align}
    t_l^{m-1}&=\gamma\sum_{k=1}^K(-1)^{b_{k,m}^m+\frac12\beta_{k,m}^m} {\rm e}^{-j\Delta_k}\|\bs h_k\|^2\sum_{n=1}^{2^{m-1}}(-1)^{\left(\bs \eta_k^{m}+\bs a_{l-1}^{m-1}\right)^{\rm T}{\bs a_{n-1}^{m-1}}}\notag\\
    &+\sum_{n=1}^{2^{m-1}}(-1)^{(\bs a_{l-1}^{m-1})^{\rm T}{\bs a_{n-1}^{m-1}}}\tilde {Z}_{n}^{m-1}\label{hadaym1}.
\end{align}
Equation \eqref{hadaym1} indicates that, if we have $\bs \eta_k^m=\bs a_{l-1}^{m-1}$, peaks will appear at frequency $\bs\eta^m_k, k\in\{1,2,\cdots,{K_i}\}$, where the maximum value is ${\rm e}^{-j\Delta_k}2^{m-1}\gamma |\bs h_k|^2$.
On this basis, $ \bs{\hat\eta}_k^m$ can be recovered by searching the largest absolute value of $\bs t^{m-1}$ and $V_{k,n}^{m-1}$ can be estimated based on $\bs{\hat\eta}_k^m$.

Furthermore, since $b_{k,m}^m=\beta_{k,m}^m=0$, the delay of device $k$ can be recovered by the phase angle of the maximum value.
\begin{align}
    {\hat\Delta}_{k,1}^m=-{\rm Arg}(\max \bs t^{m-1} )\label{Delta1}.
\end{align}

\subsubsection{Estimation of $(\bs \eta_k^{m-1},b_{k,m-1}^m,\beta_{k,m-1}^m, {\Delta}_{k,2}^m)$}\label{sectionalpham1}
After recovering $(\bs{\hat\eta}_k^{m},{\hat \Delta}_{k,1}^m)$, we next estimate $(\bs\eta_k^{m-1},b_{k,m-1}^m,\beta_{k,m-1}^m, {\Delta}_{k,2}^m)$ in a similar way.
Define
\begin{align}\label{ym1next}
    \bs Y_i^{m-1}(n)=\frac12\left({{\rm e}^{-j\hat{\Delta}_{k,1}^m}\bs Y_{i}^m(2n-1)}+({\hat V}_{k,n}^{m-1})^{*} \bs Y_{i}^m(2n)\right).
\end{align}
Under the assumption that ${\hat {\bs V}}_k^{m-1}$ and ${\hat \Delta}_{k,1}^m$ are correctly estimated, according to \eqref{y2j0} and \eqref{y2j1}, $\bs Y_i^{m-1}(n)$ is further expressed as
\begin{align}\label{yjm1}
    \bs Y_{i}^{m-1}(n)\!&=\sqrt{\gamma}\bs h_k X_{k,n}^{m-1}{\rm e}^{-2j\Delta_kn}\!+\bs A_i^{m-1}(n) +\bs Z_i^{m-1}(n),
\end{align}
where the term
\begin{align}
    \bs A^{m-1}_i(n)=\frac{\sqrt{\gamma}}2\sum_{l\ne k}\! \bs h_l X_{l,n}^{m-1}\!{\rm e}^{-2j\Delta_ln}\left({\rm e}^{-j(\hat{\Delta}_{k,1}^m-{\Delta}_l)}+(\hat V_{k,n}^{m-1})^{*}V_{l,n}^{m-1}\right)
\end{align}
consists of all {\em interferences} from other devices which are all second order RM sequences and
\begin{align}
    \bs Z_i^{m-1}(n)=\frac12\left((\hat V_{k,n}^{m-1})^{*}\bs Z_{i}^m(2n)+{\rm e}^{-j\hat{\Delta}_{k,1}^m}\bs Z_{i}^m(2n-1)\right)\sim {\cal CN}\left(\bs 0,\frac{1}{2}\bs I\right),
\end{align}
i.e., the variance of the channel noise is reduced by half.
Besides, we have
\begin{align}
    \left\|\frac12\bs h_l\left({\rm e}^{-j(\hat{\Delta}_{k,1}^m-{\Delta}_l)}+(\hat V_{k,n}^{m-1})^{*}V_{l,n}^{m-1}\right)\right\|\le \|\bs h_l\|
\end{align}
which indicates that the equivalent channel gain of the interferences is reduced.


When $n=1,\cdots,2^{m-2}$, applying Proposition.\ref{Proposition1} on \eqref{yjm1} leads to
\begin{align}
    \bs Y_{i}^{m-1}(2n)\!&=\sqrt{\gamma}\bs h_k X_{k,2n}^{m-1}{\rm e}^{-4j\Delta_kn}\!+\bs A^{m-1}_i({2n}) +\bs Z_{i}^{m-1}(2n)\\
    &=\sqrt{\gamma}\bs h_k V_{k,n}^{m-2}X_{k,n}^{m-2}{\rm e}^{-4j\Delta_kn}\!+\bs A^{m-1}_i(2n) +\bs Z_{i}^{m-1}(2n),
\end{align}
and
\begin{align}
    \bs Y_{i}^{m-1}(2n-1)&=\sqrt{\gamma}\bs h_k X_{k,2n-1}^{m-1}{\rm e}^{-2j\Delta_k(2n-1)}+\bs A^{m-1}_i(2n-1) +\bs Z_{i}^{m-1}(2n-1)\\
    &=\sqrt{\gamma}\bs h_kX_{k,n}^{m-2}{\rm e}^{-2j\Delta_k(2n-1)}+\bs A^{m-1}_i(2n-1) +\bs Z_{i}^{m-1}(2n-1),
\end{align}
Let $\tilde{Y}_n^{m-2}=[\bs Y_{i}^{m-1}(2n)]^{\rm T}[\bs Y_{i}^{m-1}(2n-1)]^{*}$, we have
\begin{align}\label{yjm21}
    \tilde{Y}_n^{m-2}&={\gamma}\|\bs h_k\|^2V_{k,n}^{m-2}{\rm e}^{-2j\Delta_k}\!+\tilde{Z}_n^{m-2}
\end{align}
where
\begin{align}
    V_{k,n}^{m-2}=(-1)^{b_{k,m-1}^m+\frac12\beta_{k,m-1}^m+(\bs \eta_{k}^{m-1})^{\rm T}\bs a_{n-1}^{m-2}}
\end{align}
and
\begin{align}
    &\tilde{Z}_n^{m-2}=(\sqrt{\gamma}\bs h_k V_{k,n}^{m-2}X_{k,n}^{m-2}{\rm e}^{-4j\Delta_kn})^{\rm T}[\bs A^{m-1}_{i}(2n-1) +\bs Z_{i}^{m-1}(2n-1)]^{*} \notag \\
    &+  [\bs A^{m-1}_{i}(2n) +\bs Z_{i}^{m-1}(2n)]^{\rm T}[\sqrt{\gamma}\bs h_kX_{k,n}^{m-2}{\rm e}^{-2j\Delta_k(2n-1)}\!+\bs A^{m-1}_{i}(2n\!-\!1) +\bs Z_{i}^{m-1}(2n\!-\!1)]^{*}.
\end{align}

Similar to \eqref{ym1}, applying WHT on $\tilde {\bs Y}^{m-2}=[\tilde{Y}_1^{m-2},\cdots,\tilde{Y}_{2^{m-2}}^{m-2}]^{\rm T}$ yields
\begin{align}
    t_l^{m-2}&=(\bs w_l^{m-2})^{\rm T}\tilde {\bs Y}^{m-2}\\
    &=\sum_{n=1}^{2^{m-2}}(-1)^{(\bs a_{l-1}^{m-2})^{\rm T}{\bs a_{n-1}^{m-2}}}\left(\gamma \|\bs h_k\|^2 V_{k,n}^{m-2}{\rm e}^{-2j\Delta_k} \!+\!  \tilde {Z}_{n}^{m-2}\right)\\
    &=\gamma(-1)^{b_{k,m-1}^m+\frac12\beta_{k,m-1}^m} {\rm e}^{-2j\Delta_k}\|\bs h_k\|^2\sum_{n=1}^{2^{m-2}}(-1)^{\left(\bs \eta_k^{m-1}+\bs a_{l-1}^{m-2}\right)^{\rm T}{\bs a_{n-1}^{m-2}}}\notag\\
    &+\sum_{n=1}^{2^{m-2}}(-1)^{(\bs a_{l-1}^{m-2})^{\rm T}{\bs a_{n-1}^{m-2}}}\tilde {Z}_{n}^{m-2}\label{hadaym2}.
\end{align}
Equation \eqref{hadaym2} indicates that $\bs {\hat\eta}_k^{m-1}$ can be recovered by searching the maximum value of the result. Comparing \eqref{ym1} and \eqref{yjm21}, we know that $\bs {\hat\eta}_k^{m-1}$ is more likely to be correctly estimated than $\bs {\hat\eta}_k^{m}$ because the variance of channel noise is reduced by half.

Moreover, we have
\begin{align}\label{bmbetam}
    (-1)^{b_{k,m-1}^m+\frac12\beta_{k,m-1}^m}=\left\{\begin{array}{rc}
                                        -i & \ {\rm if}\ (b_{k,m-1}^m,\beta_{k,m-1}^m)=(1,1), \\
                                        -1 & \ {\rm if}\ (b_{k,m-1}^m,\beta_{k,m-1}^m)=(1,0), \\
                                        i & \ {\rm if}\ (b_{k,m-1}^m,\beta_{k,m-1}^m)=(0,1), \\
                                        1 & \ {\rm if}\ (b_{k,m-1}^m,\beta_{k,m-1}^m)=(0,0).
                                      \end{array}\right.
\end{align}
Since $\Delta_k\approx \hat\Delta_{k,1}^m$, equation \eqref{bmbetam} indicates that $(b_{k,m-1}^m,\beta_{k,m-1}^m)$ can be estimated by the polarity of the largest value of ${\rm e}^{2j{\hat\Delta}_{k,1}^m}\bs t^{m-2}$.
For example, if the real part of the maximum value is positive and greater than the absolute value of the imaginary part, then we have $(b_{k,m-1}^m,\beta_{k,m-1}^m)=(0,0)$.

Further, $V_{k,n}^{m-2}$ is recovered through
\begin{equation}
\begin{split}\label{vkm1}
    {\hat V}_{k,n}^{m-2}=(-1)^{\hat b_{k,m-1}^m+\frac12\hat\beta_{k,m-1}^m+(\bs {\hat\eta}_k^{m-1})^{\rm T}\bs a_{n-1}^{m-2}},
\end{split}
\end{equation}
and
\begin{align}\label{Delta2}
    {\hat\Delta}_{k,2}^m=-{\rm Arg}\left\{(\max \bs t^{m-1})\left[(-1)^{\hat b_{k,m-1}^m+\frac12\hat\beta_{k,m-1}^m}\right]^{*} \right\}.
\end{align}


\subsubsection{Estimation of Channel Coefficient $(b_{k,1}^m,\beta_{k,1}^m,h_k)$}\label{sectionalpham11}
We continue these process until all the estimates $\left(\bs {\hat\eta}_k^s, {\hat b}_{k,s}^m, {\hat\beta}_{k,s}^m\right), s\in\{2,\cdots,m\}$ and $\hat{\Delta}_{k,m-1}^m$ are obtained.
We have ${\rm e}^{j\hat{\Delta}_{k,m-1}^m} \approx {\rm e}^{2^{m-2}j{\Delta}_{k}}$.

According to \eqref{yjm1}, the received sequence in the last layer can be written as
\begin{align}
    \bs Y_{i}^{1}(n)&=\sqrt{\gamma} \bs h_k X_{k,n}^{1}{\rm e}^{-2^{m-1}j\Delta_kn}+\bs A_i^1(n)+\bs Z_i^{1}(n), \qquad n=1,2,
\end{align}
where the term $\bs A_i^1(n)$ consists of all {\em interferences} from other devices which are all second order RM sequences and $\bs Z_i^{1}(n)\sim{\cal}(\bs 0,\frac1{2^{m-1}}\bs I)$.

Accordingly, we have
\begin{align}\label{y11}
    \bs Y_{i}^{1}(1)&=\sqrt{\gamma} \bs h_k X_{k,1}^{1}{\rm e}^{-2^{m-1}j\Delta_k}+\bs A_i^1(1)+\bs Z_i^{1}(1)\\
    &=\sqrt{\gamma} \bs h_k {\rm e}^{-2^{m-1}j\Delta_k}+\bs A_i^1(1)+\bs Z_i^{1}(1)
\end{align}
and
\begin{align}\label{y21}
    \bs Y_{i}^{1}(2)&=\sqrt{\gamma} \bs h_k X_{k,2}^{1}{\rm e}^{-2^{m}j\Delta_k}+\bs A_i^1(2)+\bs Z_i^{1}(2)\\
   &=\sqrt{\gamma} \bs h_k (-1)^{b_{k,1}^m+\frac12\beta_{k,1}^m}{\rm e}^{-2^{m}j\Delta_k}+\bs A_i^1(2)+\bs Z_i^{1}(2)
\end{align}
Similar to previous processing, define $\tilde{Y}_1^0=(\bs Y_i^1(2))^{\rm T}(\bs Y_i^1(1))^{*}$, we have
\begin{align}\label{y10}
    \tilde{Y}_1^0={\gamma}|\bs h_k|^2(-1)^{b_{k,1}^m+\frac12\beta_{k,1}^m}{\rm e}^{-2^{m-1}j\Delta_k}+\tilde Z_1^0,
\end{align}
where
\begin{align}
    \tilde{Z}_1^0&=(\sqrt{\gamma} \bs h_k (-1)^{b_{k,1}^m+\frac12\beta_{k,1}^m}{\rm e}^{-2^{m}j\Delta_k}+\bs A_i^1(2)+\bs Z_i^{1}(2))^{\rm T}(\bs A_i^1(1)+\bs Z_i^{1}(1))^{*}\\
    &+(\bs A_i^1(2)+\bs Z_i^{1}(2))^{\rm T}(\sqrt{\gamma} \bs h_k {\rm e}^{-2^{m-1}j\Delta_k})^{*}
\end{align}
According to \eqref{y10}, $(b_{k,1}^m,\beta_{k,1}^m)$ can be estimated by the polarity of ${\rm e}^{2j\hat{\Delta}_{k,m-1}^m}\tilde{Y}_1^0$.
And $\hat{\Delta}_{k,m}^m$ can be estimated as
\begin{align}\label{delta2m1}
   \hat{\Delta}_{k,m}^m=-{\rm Arg}\left(\tilde{Y}_1^0\left((-1)^{\hat b_{k,1}^m+\frac12\hat\beta_{k,1}^m}\right)^{*} \right).
\end{align}
Moreover, the channel coefficient $\bs h_k$ can be estimated as
\begin{align}\label{hk}
    \sqrt{\gamma}\hat {\bs h}_k=\frac1{2}\left({\bs Y_i^1(1){\rm e}^{j\hat{\Delta}_{k,m}^m} +\left((-1)^{\hat b_{k,1}^m+\frac12\hat\beta_{k,1}^m}\right)^{*}{\rm e}^{2j\hat{\Delta}_{k,m}^m}\bs Y_i^1(2)}\right).
\end{align}
Note that we can infer $\hat {\bs h}_k$ from \eqref{hk} if the AP knows the value of $\gamma$, but we only need to know the product of the two.

\subsubsection{Estimation of $\Delta_k$}\label{sectiondeltak}
So far, the matrix-vector pair $\left(\hat{\bs P}_k^m,\hat{\bs b}_k^m\right)$, the corresponding channel coefficient ${\hat{\bs h}}_k$, and $ {\bs{\hat\Delta}}_{k}^m$ has been completely estimated. In addition, $\hat{\bs X}_k^m$ can be obtained through \eqref{cm2}.
To estimate the remaining devices, however, we still need to estimate $\Delta_k$ to remove the interference of device $k$ according to \eqref{OFDMAPdiscreteDFTSIC}.

Let $\hat\Delta_{k,l}^m=\Delta_{k,l}^m+\delta_{k,l}^m, l=1,\cdots,m$, where $\delta_{k,l}^m$ indicates the estimated error.
Typically, $ \{{\delta}_{k,l}^m\}, l=1,\cdots,m$ is a small number.
Thus ${\hat\Delta}_k$ can be estimated through ${\hat\Delta}_{k,1}^m$ directly, i.e., ${\hat\Delta}_k={\hat\Delta}_{k,1}^m$.
Then, to remove the interference of device $k$ according to \eqref{OFDMAPdiscreteDFTSIC}, a straightforward way of evaluating the phase angle $n\hat\Delta_k, n=1,\cdots,N,$ is
\begin{align}
    {\rm Arg}({\rm e}^{ jn \hat\Delta_{k,1}^m })={\rm Arg}({\rm e}^{ jn \delta_{k,1}^m }{\rm e}^{ jn \Delta_{k,1}^m }).\label{delayerror}
\end{align}
However, in this way, the error of the phase angle between $n\hat\Delta_k$ and $n\Delta_k$ becomes $n \delta_{k,1}^m$, which will become large for large $n$.

Since we know the error $ \{{\delta}_{k,l}^m\}, l=1,\cdots,m$ is a small number. We can estimate $\hat\Delta_k$ through $\bs{{\hat\Delta}}_k^m$ instead of just $\hat\Delta_{k,1}^m$.
Let $\bs{{\tilde\Delta}}_k^m=\left[{{\tilde\Delta}}_{k,1}^m,\cdots,{{\tilde\Delta}}_{k,m}^m\right]$ where
\begin{align}
    {{\tilde\Delta}}_{k,l}^m={\rm Arg}\left({\rm e}^{j2^{l-1}\left({\hat\Delta}_{k,1}^m-e_{k}^m\right)} \right),\quad l=1,\cdots,m.
\end{align}
Here $e_{k}^m$ is expected to be the estimation error $\delta_{k,1}^m$ that takes values around 0, and can be evaluated as
\begin{align}
    e_{k}^m=\argmin\left\{ \left\|{\hat{{\bs\Delta}}_{k}^m}-{\tilde{{\bs\Delta}}_{k}^m}  \right\|\right\}.
\end{align}.
Then $\hat\Delta_k$ is estimated as
\begin{align}
    \hat\Delta_k&=\hat\Delta_{k,1}^m-e_{k}^m.\label{delayerror}
\end{align}

\subsubsection{The {\bf findPb} Algorithm}

According to the analysis from Section \ref{sectionalpham} to Section \ref{sectiondeltak}, the detailed algorithm is summarized as in Algorithm 2.

\begin{center}\begin{tabular}{l l}
\toprule
\multicolumn{1}{l}{{\bf Algorithm 2}: {\bf findPb} function.}\\
\midrule
{\bf Input}: the received signal $\bs Y^m_i$, the maximum number of active devices $K_{\rm max}$.\\
{\bf while} $\|\bs Y^q_i\|_F>\varepsilon $\footnotemark[1] and $k<K_{\rm max}$\footnotemark[2] {\bf do}\\
\quad $k\leftarrow k+1$.\\
\quad {\bf for }$s=m,m-1,\cdots,2$ {\bf do}\\
\qquad Split $\bs Y^s_i$ into two partial sequences similar to \eqref{y2j0} and \eqref{y2j1}.\\
\qquad Perform the element-wise conjugate multiplication according to \eqref{ym1}.\\
\qquad Perform WHT and recover $\hat{\bs \eta}^s$ by the binary index of the largest component.\\
\qquad {\bf if} $s=m$ {\bf do}\\
\qquad\quad Set $\hat b_{k,m}^m=0,\hat \beta_{k,m}^m=0$ and recover $\Delta_{k,1}^m$ according to \eqref{Delta1}.\\
\qquad {\bf else} {\bf do}\\
\qquad\quad Recover $(\hat b_{k,s}^m,\hat \beta_{k,s}^m)$, $\hat{\bs v}_k^{s-1}$, and $\Delta_{k,2}^m$ according to \eqref{bmbetam} -- \eqref{Delta2}, respectively.\\
\qquad {\bf end if}\\
\qquad Calculate $\bs Y^{s-1}_i$ according to \eqref{ym1next}.\\
\quad {\bf end for}\\
\quad Recover $(\hat b_{k,1}^m,\hat \beta_{k,1}^m)$ according to \eqref{bmbetam} and \eqref{y10}.\\
\quad Add $(\hat{\bs P}_k^m, \hat{\bs b}_k^m)$ to the decoded set.\\
\quad Calculate the codeword $\hat{\bs X}_k^m$ according to $(\hat{\bs P}_k^m, \hat{\bs b}_k^m)$ and estimate $\sqrt{\gamma}\hat {\bs h}_k$ according to \eqref{hk}.\\
\quad Recover $\hat\Delta_{k}$ according to \eqref{delayerror}.\\
\quad Remove the interference of device $k$ and update $\bs Y^m_i$ according to \eqref{OFDMAPdiscreteDFTSIC}.\\
\quad Break and delete the detected message if $\|\bs Y^m_i\|_F$ becomes larger after removing device $k$.\\
{\bf end while}\\
{\bf Output}: $\left(\hat{\bs P}_l^m, \hat{\bs b}_l^m\right)$, ${\hat{\Delta}}_l$, and $\hat {\bs h}_l$ for $l=1,2,\cdots, k$.\\
\bottomrule
\end{tabular}
\footnotetext[1]{Various criteria are possible here. We found $\varepsilon=(22K^{-\frac13}r^{-\frac14}(\sigma^2+r2^q))^{\frac12}-d+3(q-p)$ works well. We should emphasize that it's very much an open question as to how to choose this parameter optimally. Our suggestion would be to try to optimize the value empirically for the regime of interest.}
\footnotetext[2]{In practice, we take the iteration limits $K_{\rm max}$ to be more than three times the expected number of messages.}
\end{center}

\section{Computational Complexity Analysis}\label{CCA}
In Algorithm 1, there are $2^d$ sub-blocks. In each sub-block, there are $2^p$ time slots, each of which is length $2^m$.
Thus the total number of iterations in Algorithm 1 is $2^{p+d}$.
In each iteration, Algorithm 1 calls Algorithm 2 to return all the messages that transmitted in the given sub-block and slots.
And the computational complexity of Algorithm 1 mainly comes from the calling of Algorithm 2 in each iteration.

For Algorithm 2, it iterates at most $K_{\rm max}$ times to obtain $K_{\rm max}$ messages. The most expensive operations in each iteration come from two parts: matrix-vector pair estimation through fast WHT and the generation of the RM code through \eqref{cm2}.
We consider the number of multiplication operations required to run the algorithm, where the number of addition operations is similar to it.
For matrix-vector pair estimation, the number of multiplication operations required in each iteration is on the order of $\sum_{s=2}^m{\cal O}((s-1+r)2^{s-1})={\cal O}((m+r-2)2^m)$.
Similarly, in each iteration, the complexity of generating the Reed-Muller code when a matrix-vector pair $\left(\hat{\bs P}^m, \hat{\bs b}^m\right)$ is found is ${\cal O}((m^2+2m)2^m)$.
In summary, the worst-case complexity of Algorithm 2 is on the order of ${\cal O}\left\{K_{\rm max}2^m\left(m^2+3m+r-2\right)\right\}$.

Since the number of sub-blocks is $2^d$ and the number of slots in each sub-block is $2^p$, the complexity of Algorithm 1 is thus ${\cal O}\left\{2^{d+p} K_{\rm max}2^m\left[m^2+3m+r-2\right]\right\}$.
In practice, the maximum number of neighbors of the AP in each slot in Algorithm 2 is set to $ K_{\rm max}=\left\lceil\frac{6K^{*}}{2^{p}}r^{\frac14}\right\rceil$.
Accordingly, the complexity of Algorithm 1 is ${\cal O}\left\{6K^{*}2^{d+m}r^{\frac14}\left(m^2+3m+r-2\right)\right\}$.

However, we should emphasize that the above analysis of complexity is the worst case.
Actually, the number of iterations of Algorithm 2 is mainly determined by $\varepsilon$, not $K_{\rm max}$.
If the active devices transmitted in the given slot is decoded, it is likely that the energy of the received signal is smaller than $\varepsilon^2$. Besides, since each active device transmits randomly, the number of active devices in each slot is three times smaller than $K_{\rm max}$ on average.
This also indicates that our algorithm does not need to know the number of active devices in each slot; furthermore, there is no need to know the total number of active devices.

\section{Numerical Results}\label{Numericalresults}

\subsection{Definition of Error Metrics}
We first define the false alarm rate and the miss rate. These are our main performance metrics.

\begin{figure}[H]
  \begin{center}
  \includegraphics[width=3in]{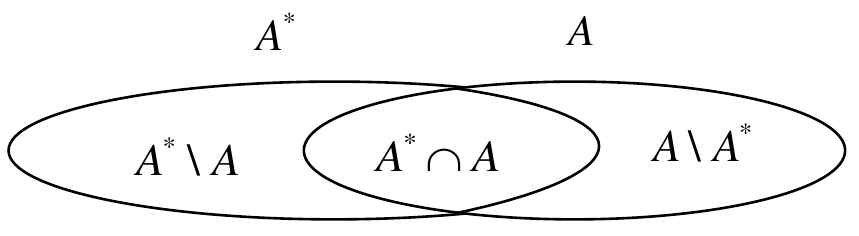}
  \end{center}
  \caption{The set relationship.}\label{setrelation}
\end{figure}

Denote $A^{*}\subset\{1,2,\cdots,2^B\}$ as the index of the messages transmitted by in-cell devices of the AP. We have $|A^{*}|=K^{*}$. And let $A\subset\{1,2,\cdots,2^B\}$ denotes the index of the output messages of the algorithm.
The set relationship is depicted in Fig. \ref{setrelation}.

In our algorithm, the AP does not know the number of its in-cell devices $K^{*}$. In this case, the algorithm output all the detected messages in $A$.
Accordingly, we define the false alarm rate in this phase as
\begin{equation}
\begin{split}\label{DefiniFAR}
    \frac{|A\backslash A^{*}|}{|A|},
\end{split}
\end{equation}
and the miss rate as
\begin{equation}
\begin{split}\label{DefiniMAR}
    \frac{|A^{*}\backslash A|}{|A^{*}|}.
\end{split}
\end{equation}
The false alarm rate and miss rate reflect the error performance of the algorithm when the number of devices  in the AP is unknown.

\subsection{Simulation Parameter Setting}

The detailed simulation parameters are listed in Table \ref{tablesimulationparameter}.
We consider a $500\times500$ ${\rm m}^2$ rectangle where the devices are randomly distributed in the plane.
According to \cite{IMT}, the number of devices on the plane with area 0.25 ${\rm km}^2$ can be up to $|\Phi|=2.5\times10^5$, however, only a small part of them are active.
In this paper, we consider the number of active devices $K$ ranges from 1000 to 8000, i.e., $K\in[1000,8000]$.

According to \eqref{averageneighbor}, we have $K^{*}\approx 1.11\times10^{-2}K$ when $r=1$.
And $K^{*}\approx1.25\times10^{-2}K$ when $r=16$.
Assuming $K=1000$, we have $K^{*}\approx11$ when $r=1$ and $K^{*}\approx13$ when $r=16$.
Assuming $K=8000$, we have $K^{*}\approx89$ when $r=1$ and $K^{*}\approx100$ when $r=16$.
The AP aims to decode the messages transmitted by its neighbor, while the messages transmitted by its non-neighbor are treated as interference.

\begin{table}[htbp]
\caption{Simulation Parameters.}
\centering
\begin{tabular}{l|c}
  \hline
  \label{tablesimulationparameter}
  Parameters & Value \\
  \hline
  Channel gain threshold $\theta$ & $10^{-6}$ \\
  Path-loss exponent $\alpha$ & $4$ \\
  Transmit SNR of each device $\gamma$ & 60 dB \\
  Average number of detected devices in each slot $K_{\rm max}$ & $\left\lceil\frac{6K^{*}}{2^{p}}r^{\frac14}\right\rceil$ \\
  The square area of the device distribution region $S$ & $500 \times 500\ {\rm m}^2$\\
  Carrier spacing $\Delta f$ & 15 kHz\\
  Maximum device delay $\tau_{\rm max}$ & 10 $\upmu$s\\
  \hline
\end{tabular}
\end{table}

\subsection{Synchronous Transmission}

In this subsection, we set $\tau_{\rm max}=0$, i.e., the transmissions of different devices are synchronized.
In this case, a state-of-the-art algorithm using RM codes is the list RM\_LLD algorithm given in \cite{Hanzo}.
In \cite{Hanzo}, the author uses only the subset of the RM codewords; thus encodes fewer bits, namely $B=\frac12(d+m+p)(d+m+p+1)$.
Note that the algorithm proposed in \cite{Hanzo} applies only for single antenna case. In our algorithm, we set $r=1,4,16$, respectively.
For fair comparison, we set the codelength to be $C=2^{12}=4,096$ in both algorithms.
In this case, the number of information bits in \cite{Hanzo} is $B=\frac12\times12\times(12-1)=66$ bits.
To make the transmit bits comparable, we set $d=0, m=10, p=2$ in our scheme.
Since the number of slots is small, in our scheme, each messages randomly chooses one of the 4 slots for transmitting.
Accordingly, the number of transmit bits of Algorithm 1 is $B=\frac12m(m+3)+p=67$ bits, which is comparable with \cite{Hanzo}.

\begin{figure*}[htbp]
\centering

\subfigure[]{
\begin{minipage}[t]{0.45\linewidth}
\centering
\includegraphics[width=3in]{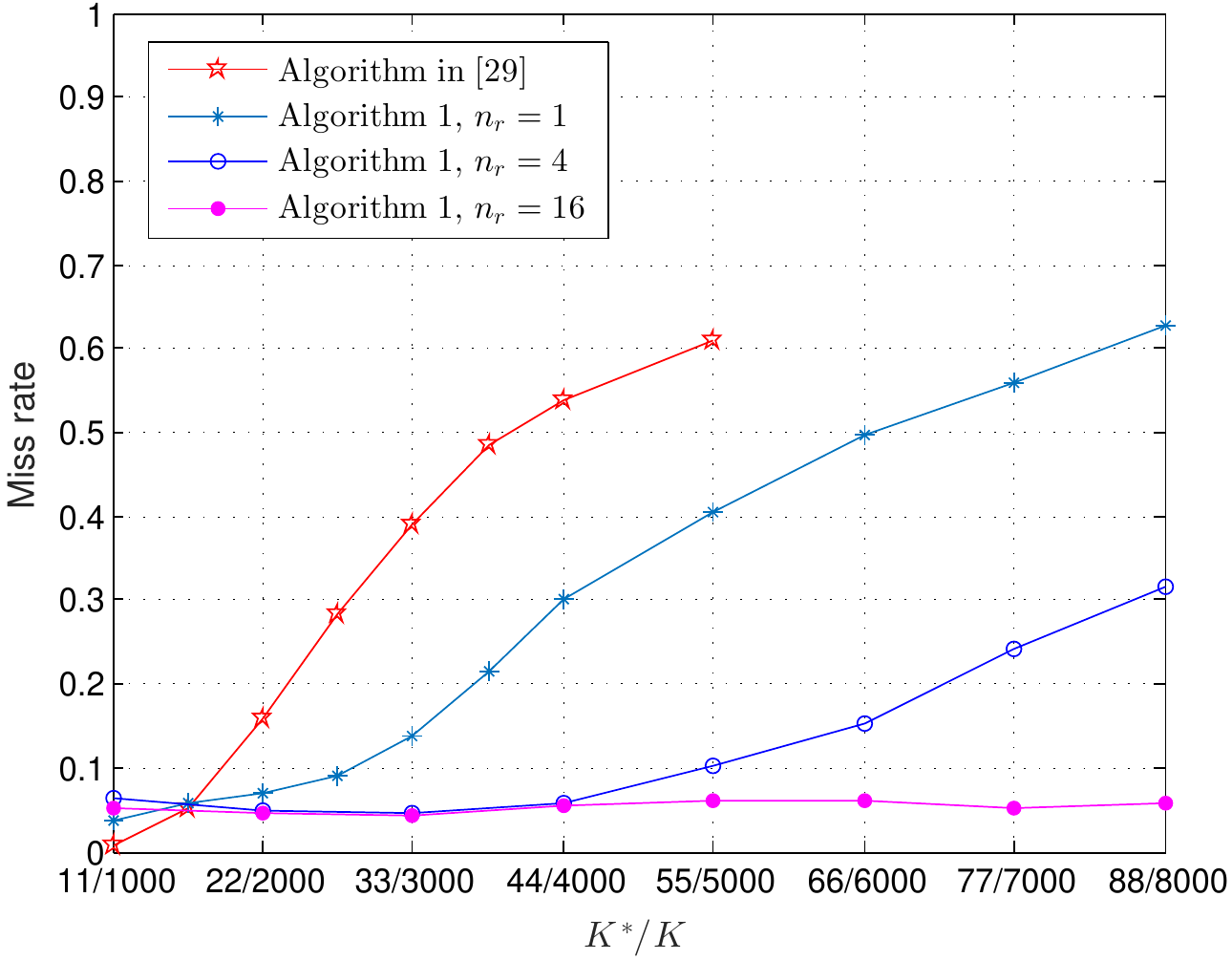}
\end{minipage}%
\label{syncSDP}
}%
\subfigure[]{
\begin{minipage}[t]{0.45\linewidth}
\centering
\includegraphics[width=3in]{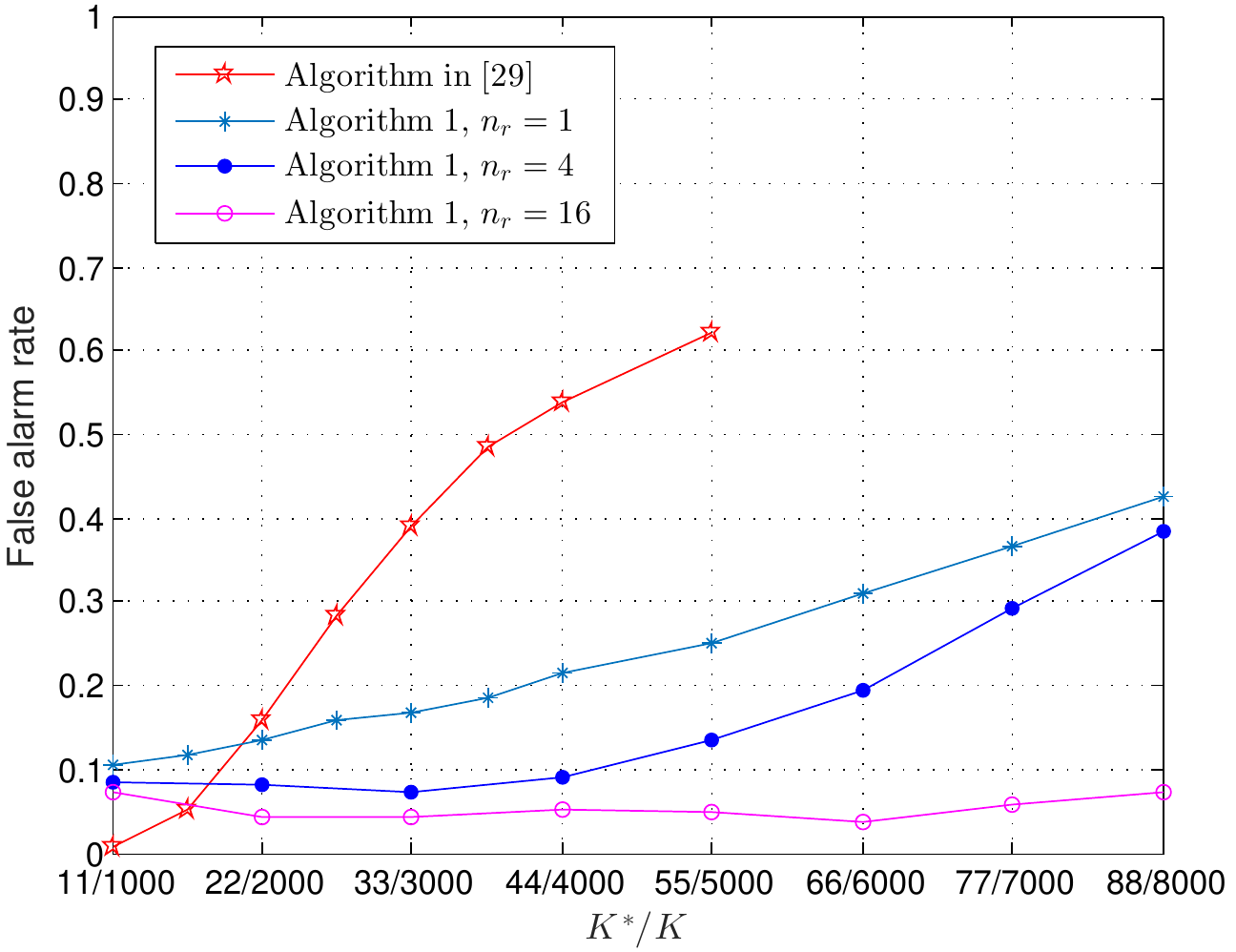}
\end{minipage}%
\label{syncFAR}
}%

\centering
\caption{Performance comparison with the algorithm in \cite{Hanzo}. (a) The miss rate and (b) the false alarm rate.}
\label{sync}
\end{figure*}

Fig. \ref{sync} illustrates the miss rate and false alarm rate of both algorithms.
In Fig. \ref{sync}, the horizontal axis is the number of neighbors of the AP $K^*$ and the total number of active devices $K$ when $r=1$, respectively.
The author in \cite{Hanzo} assume that the AP knows the number of its neighbors, thus the miss rate equals to the false alarm rate for the algorithm in \cite{Hanzo}.
In  our algorithm, the AP does not need to know the number of its neighbors.
Fig. \ref{sync} shows that the performance of Algorithm 1 improves as the number of receiving antennas increases.
Fig. \ref{sync} demonstrates that Algorithm 1 outperforms the list RM\_LLD algorithm when the number of active devices is larger than 2000. But the performance of the list RM\_LLD algorithm is better when the number of active devices is smaller than 2000, i.e., the number of neighbors of the AP is less than 22.
This is because the codelength in our algorithm is divided into 4 slots to reduce the number of active devices in each slot. However, when the number of active users is small (for example, below 2000), there is no need to use slotting.
This suggests us that the number of slots should be reduced if the number of active devices is small.
Fortunately, our algorithm can flexibly change the number of slots to deal with different situations.


\subsection{Asynchronous Transmission}
In this subsection, we study the performance of Algorithm 1 under asynchronous transmission.
The maximum delay is set to be $\tau_{\rm max}=10\ \upmu$s, further, the length of the cyclic prefix can be calculated as $M=\left\lceil\frac{3}{20}2^m\right\rceil$.
We assume the normalized transmission delay $\Delta_k=2\pi\Delta f\tau_k$ is uniformly distributed in $[-\pi,\pi]$.
As far as we know, this paper is the first using RM codes to handle continuous transmission delay in massive access.

\begin{figure*}[htbp]
\centering

\subfigure[]{
\begin{minipage}[t]{0.45\linewidth}
\centering
\includegraphics[width=3in]{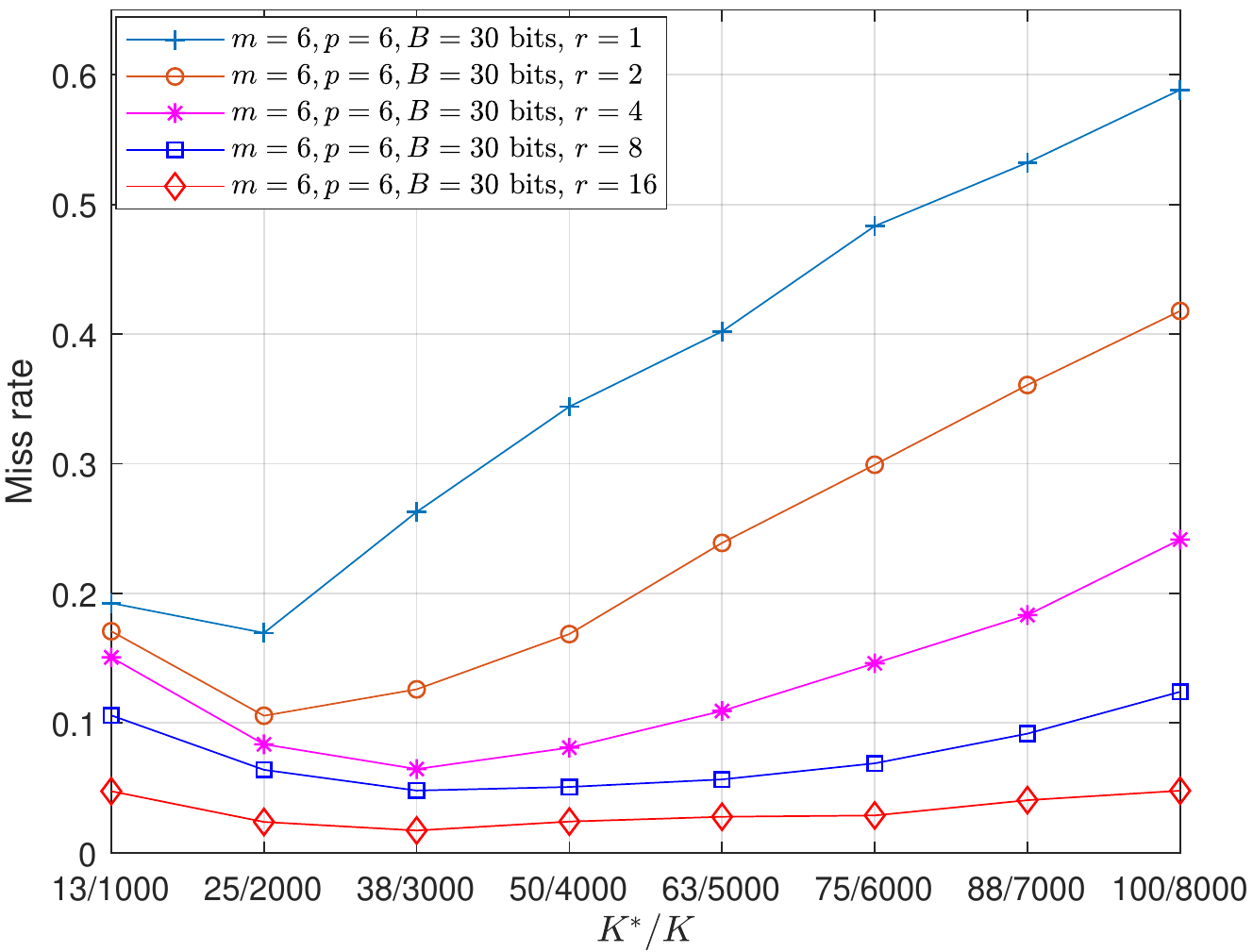}
\end{minipage}%
\label{SDP}
}%
\subfigure[]{
\begin{minipage}[t]{0.45\linewidth}
\centering
\includegraphics[width=3in]{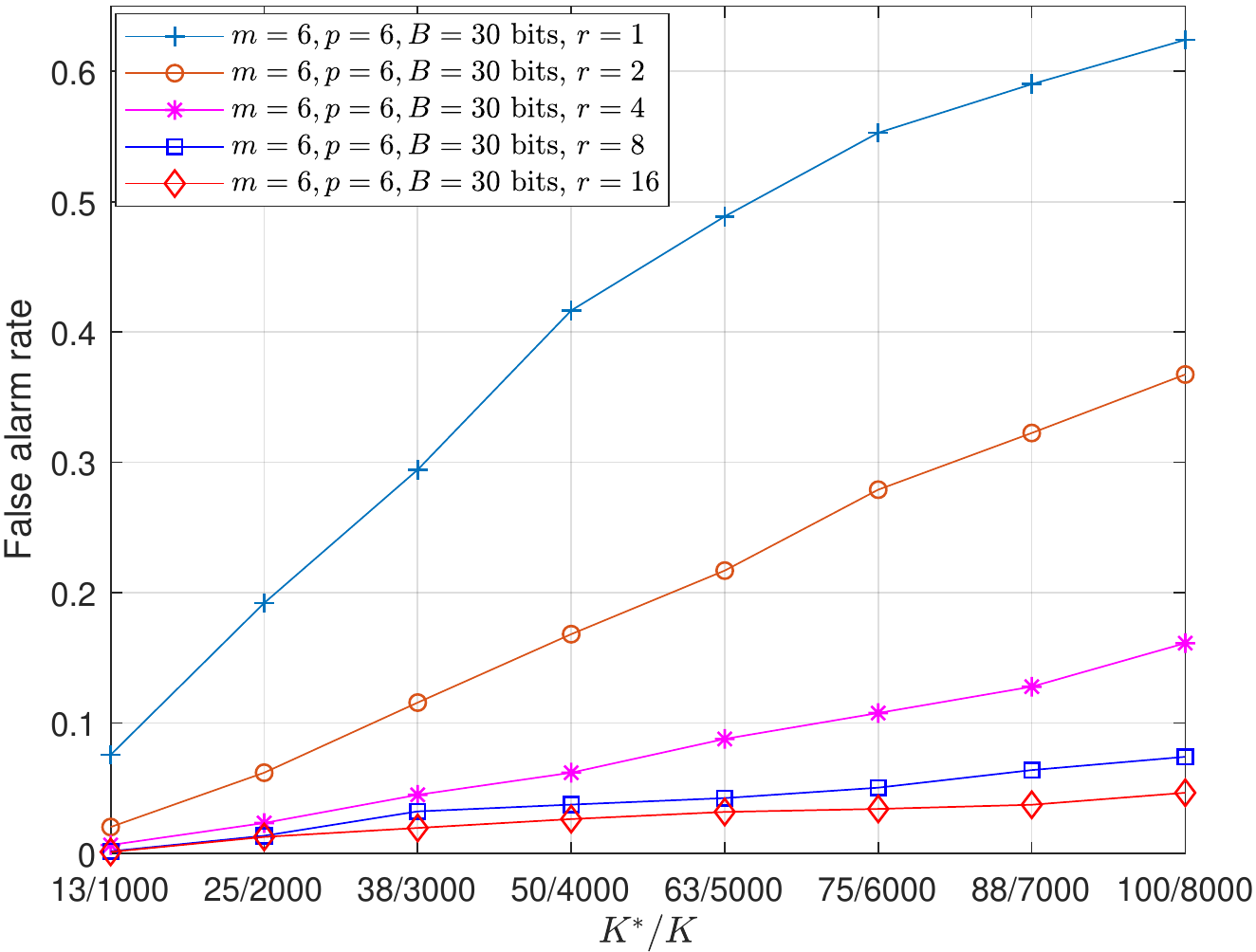}
\end{minipage}%
\label{FAR}
}%

\centering
\caption{Performance comparison with different number of receiving antennas. (a) The miss rate and (b) the false alarm rate.}
\label{fig3}
\end{figure*}

Fig. \ref{fig3} depicts the miss rate and false alarm rate obtained by Algorithm 1 with different number of receive antennas.
We set $q=6, p=6, d=0$, which means that the number of slots and the length of each slot are both $2^6=64$ and $M=10$.
In Fig. \ref{fig3}, the horizontal axis is $K^*$ and $K$ when $r=16$, respectively.
In this case, the total number of information bits can be transmitted is $30$ bits according to \eqref{B}, and the codelength is $2^6(2^6+10)=4,736$ according to \eqref{Codelength}.

Fig. \ref{fig3} shows that the performance of Algorithm 1 improves as the number of receiving antennas increases.
If the number of receive antennas is increased from $r=1$ to $r=2$, the performance of Algorithm 1 will be greatly improved.
However, if we further increase the number of receiving antennas, the increase of the performance is limited.
Differently, we observe in Fig. \ref{fig3} that the miss rate at $K=1000$ is greater than the miss rate at $K=2000$, which might appear surprising.
We believe that this behavior is due to the suboptimal tuning of the parameters $\varepsilon$ and $ K_{\rm max}$ in Algorithm 1.
One of the advantage of Algorithm 1 is that no tuning of parameters is required.
We use the same parameter choices throughout the paper.
It is worth notice that both the miss rate and false alarm rate is below 0.05 when $r=16$.
This indicates that we need use multiple antennas for practice use.

\begin{figure*}[htbp]
\centering

\subfigure[]{
\begin{minipage}[t]{0.45\linewidth}
\centering
\includegraphics[width=3in]{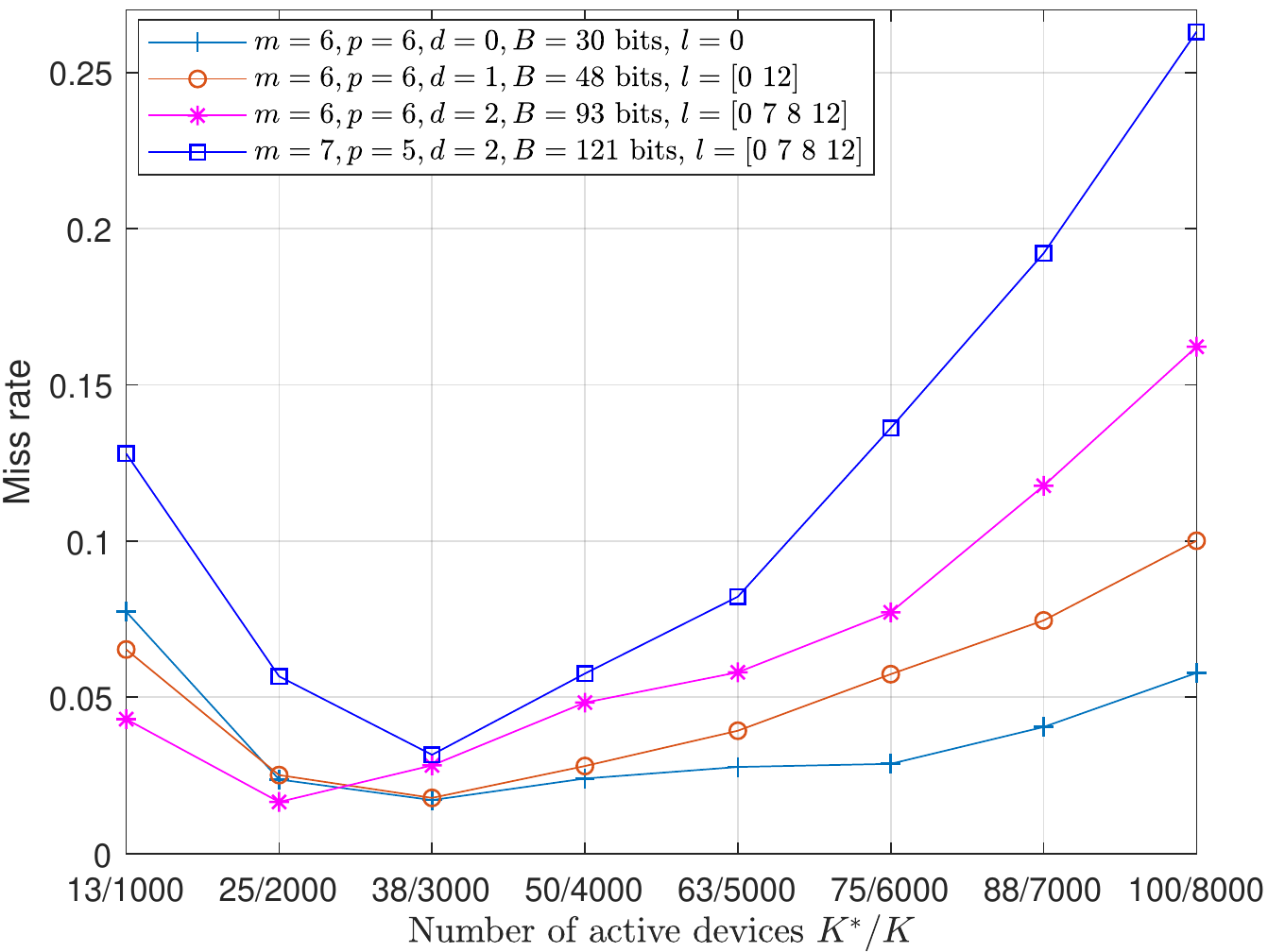}
\end{minipage}%
\label{SDP}
}%
\subfigure[]{
\begin{minipage}[t]{0.45\linewidth}
\centering
\includegraphics[width=3in]{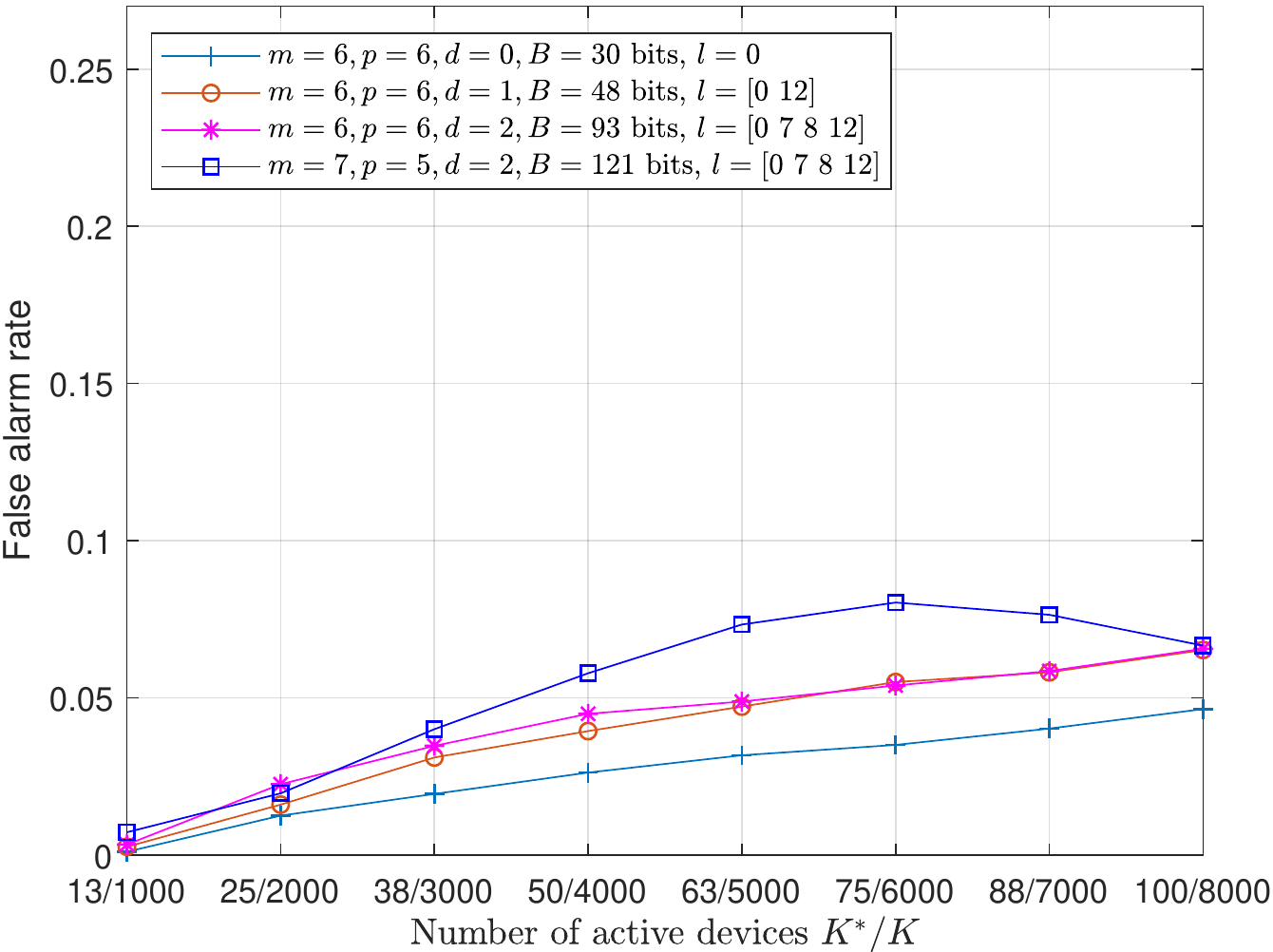}
\end{minipage}%
\label{FAR}
}%

\centering
\caption{Performance comparison with different number of receiving antennas. (a) The miss rate and (b) the false alarm rate.}
\label{fig4}
\end{figure*}

Fig. \ref{fig4} shows the miss rate and false alarm rate versus the number of active devices with different number of information bits.
The number of receive antennas is $r=16$ and $m+p=12$.
For $d=0, 1, 2$, the number of sub-blocks is $1, 2, 4$, respectively, and the corresponding codelengths are $C=4,736, C=9,472$, and $C=18,944$, respectively.
To patch the sub-blocks together, we add different number of parity check bits for different number of sub-blocks.
These choices are made with a view of minimizing the number of bits devoted to parity check while keeping the probability of information loss in the patching process pretty low.
We should emphasize that these choice are not optimal.
Generally, the miss rate and the false alarm rate are related to the parity check bits.
The more parity check bits, the worse the miss rate and the better the false alarm rate.
We refer the reader to \cite{Amalladinne2} for details of this method.
Normally, the more information bits we transmit, the worse the error performance.
For example, when the codelength is $18,944$, the error performance of the curve with $B=93$ bits is better than the curve with $B=121$ bits.
Besides, the error performance of the curve with $B=30$ bits and $B=48$ bits are comparable, but the error performance of the curve with $B=93$ bits is much worse than that of the curve with $B=30$ bits and $B=48$ bits.
This indicates two things: 1) because we need to append some parity check bits, even though we double the number of sub-blocks, the number of information bits will not double; 2) we find that only a small number of sub-blocks (up to 4) is beneficial. The error performance will be much worse if more than 4 sub-blocks are adopted, which is somewhat deviated from the implementation of using 11 sub-blocks in \cite{Amalladinne12}.

\section{Conclusion}\label{conclusion}
This paper has developed a new technique for asynchronous massive access using OFDM signaling and RM codes. The access points are assumed to have a large number of antennas. The proposed technique allows a flexible number of information bits and codelength.
Numerical results demonstrate the effectiveness of the proposed technique as well as the gains due to a large number of antennas at the access points.


\end{document}